\documentclass[sigconf]{acmart}
\usepackage{multirow}
\usepackage{colortbl}
\usepackage{dcolumn}

\AtBeginDocument{%
  \providecommand\BibTeX{{%
    \normalfont B\kern-0.5em{\scshape i\kern-0.25em b}\kern-0.8em\TeX}}}

\copyrightyear{2024} 
\acmYear{2024} 
\setcopyright{rightsretained} 
\acmConference[SIGIR '24]{Proceedings of the 47th International ACM SIGIR Conference on Research and Development in Information Retrieval}{July 14--18, 2024}{Washington, DC, USA}
\acmBooktitle{Proceedings of the 47th International ACM SIGIR Conference on Research and Development in Information Retrieval (SIGIR '24), July 14--18, 2024, Washington, DC, USA}
\acmDOI{10.1145/3626772.3657796}
\acmISBN{979-8-4007-0431-4/24/07}

\begin{document}

\title{Evaluating Search System Explainability with Psychometrics and Crowdsourcing}

\author{Catherine Chen}
\email{catherine_s_chen@brown.edu}
\affiliation{
  \institution{Brown University}
  \city{Providence}
  \state{Rhode Island}
  \country{USA}
}

\author{Carsten Eickhoff}
\email{carsten.eickhoff@uni-tuebingen.de}
\affiliation{
  \institution{University of T\"{u}bingen}
  \city{T\"{u}bingen}
  \country{Germany}
}

\renewcommand{\shortauthors}{Catherine Chen and Carsten Eickhoff}

\begin{abstract}
As information retrieval (IR) systems, such as search engines and conversational agents, become ubiquitous in various domains, the need for transparent and explainable systems grows to ensure accountability, fairness, and unbiased results. Despite recent advances in explainable AI and IR techniques, there is no consensus on the definition of explainability. Existing approaches often treat it as a singular notion, disregarding the multidimensional definition postulated in the literature. In this paper, we use psychometrics and crowdsourcing to identify human-centered factors of explainability in Web search systems and introduce SSE (Search System Explainability), an evaluation metric for explainable IR (XIR) search systems. In a crowdsourced user study, we demonstrate SSE's ability to distinguish between explainable and non-explainable systems, showing that systems with higher scores indeed indicate greater interpretability. We hope that aside from these concrete contributions to XIR, this line of work will serve as a blueprint for similar explainability evaluation efforts in other domains of machine learning and natural language processing. 
\end{abstract}



\maketitle

\section{Introduction}

Explainable information retrieval (XIR) research aims to develop methods that increase the transparency and reliability of information retrieval systems. XIR systems are designed to provide end-users with a deeper understanding of the rationale underlying ranking decisions. Besides casual Web search, these systems hold promising potential for impactful real-world information needs, such as matching patients to clinical trials, retrieving case law for legal research, and detecting misinformation in news and media analysis. Despite several advancements in their development, there is a lack of empirical, standardized techniques for evaluating the efficacy of XIR systems. 

Current approaches toward evaluating XIR systems are limited by a lack of consensus in the broader explainable artificial intelligence (XAI) community on a definition of explainability. Explainability has often been treated as a monolithic concept although recent literature suggests it to be an amalgamation of several sub-factors \citep{lipton2018mythos, doshi2018considerations, nauta2022anecdotal}. As a result, evaluation has occurred on a binary scale, considering systems as either explainable or black-box, hindering direct system comparison. Furthermore, explainability is often declared with anecdotal evidence rather than measured quantitatively. 
To address these shortcomings, we (1) identify individual factors of explainability and integrate them into a multidimensional model, and (2) provide a continuous-scale evaluation metric for explainable search systems. 

Inspired by previous work on multidimensional relevance modeling~\citep{zhang2014multidimensional}, we leverage \textit{psychometrics} \citep{furr2021psychometrics} and crowdsourcing to do so. Psychometrics is a well-established field in psychology used to develop measurement models for cognitive constructs that cannot be directly measured. Our approach involves several phases. First, we identified an exhaustive list of well-discussed explainability aspects in the community to quantitatively test. Next, we designed a user study to confirm a multidimensional model, utilizing crowdsourcing as a data-driven and efficient means of collecting diverse results from laypeople, consistent with the assumption of Web search not requiring domain-specific knowledge. Finally, we used the outcomes from our crowdsourced study to establish a metric via structural equation modeling.

This paper empowers users to understand the search systems that cater to their daily information needs in an environment potentially fraught with biases and misinformation. Specifically we contribute the following:
\begin{itemize}
    \item Leverage psychometrics and crowdsourcing to test well-discussed aspects of explainable Web search systems from the literature
    \item Introduce \textit{SSE}\footnote{Pronounced `es-es-e'}, a quantitative evaluation metric for measuring \textit{Search System Explainability} on a continuous scale
    \item Conduct a crowdsourced user study to validate SSE, gain practical insights into implementing human-centered evaluation tools, and assess the impact such tools have on human annotators
\end{itemize}

The remainder of this paper is structured as follows: Section \ref{related_work} presents background and related work on psychometric studies, the multidimensionality of explainability, and previous attempts to evaluate explainable systems. Section \ref{study_design} outlines steps taken to develop our measuring instrument and crowdsourcing task setup. Section \ref{data_analysis} presents the results of our data collection and model creation efforts. Section \ref{metric} proposes SSE and examines its effectiveness in evaluating search system explainability. Finally, Section \ref{discussion} analyzes the dimensions of explainability users found important and discusses implications for future XIR system design and evaluation. Section~\ref{conclusion} concludes with an overview of future work.

\section{Related Work} \label{related_work}

\subsection{Psychometrics, SEM, and Crowdsourcing}

Psychometrics uses Structural Equation Modeling (SEM) to construct models from observed data by measuring the presence of latent variables (factors) through observed variables (questionnaire items) \citep{ullman2012structural, furr2021psychometrics}. SEM consists of two parts: (1) \textit{Exploratory Factor Analysis (EFA)} to produce a hypothesized model structure and (2) \textit{Confirmatory Factor Analysis (CFA)} to confirm the EFA-derived model fit on a held-out dataset. EFA identifies the number of latent factors and which items load on the discovered dimensions using a statistical technique that iteratively groups and prunes items to reach a high-quality estimate of covariance in the observed data set. CFA re-estimates model parameters using maximum likelihood on a held-out set of observed data and assesses model fit via statistical significance testing of multiple alternative models.

Since SEM requires large amounts of user response data, crowdsourcing is often used for data collection due to its convenience and efficiency in quickly recruiting a large number of participants. 
However, ensuring data quality in crowdsourcing is challenging, since the payout may be the main motivator for workers to complete tasks and platforms become more saturated with low-quality workers. To mitigate these issues, preventative measures can be taken by setting high worker qualifications, enabling rigorous quality control checks, and post-processing data for inattentive responses to verify quality work \citep{johnson2005ascertaining, behrend2011viability, meade2012identifying, goldberg1985prediction}. We describe the quality control checks we employ in our study in Section \ref{study_design}.

\subsection{Evaluation of Explainable Systems}

Explainability is still often considered to be a binary concept despite recent literature that suggests that it may be best measured as a combination of several factors \citep{doshi2018considerations, lipton2018mythos, nauta2022anecdotal}. \citet{lipton2018mythos} and \citet{doshi2017towards} suggest that explainability is (1) ill-defined with no consensus and (2) an amalgamation of several factors rather than a monolithic concept. Specifically, both recognize the need to ground the explainability in the context of certain desiderata, such as \textit{trustworthiness} or \textit{causality}. \citet{nauta2022anecdotal} additionally identify 12 such conceptual properties for the systematic, multidimensional evaluation of explainability. 

Due to the lack of consensus and recognition of explainability as a multi-faceted concept, evaluation falls short in two ways. Firstly, current methods do not consider a multidimensional definition. Authors of frameworks such as LIME and SHAP demonstrate the utility of their methods through user evaluations but fail to quantify the degree of explainability provided by their methods \citep{ribeiro2016should, lundberg2017unified}. Relying solely on declarations of explainability without measurement hinders targeted system improvements and a more holistic definition is needed for robustness. Secondly, while there are many ML evaluation metrics for system performance comparison, there is no standard metric for evaluating explainability. Several authors acknowledge the importance of quantitative evaluation metrics \citep{nauta2022anecdotal, doshi2017towards, adadi2018peeking, das2020opportunities} and while \citet{nguyen2020quantitative} come close by introducing a suite of metrics to quantify interpretability, they fail to quantify interactions between facets and measure the importance of each individual facet. 

Existing approaches for explainability in IR often focus on a singular aspect of explainability or lack evaluation of explanation quality, relying on anecdotal evidence \citep{singh2019exs, qu2020towards, yu2022towards, polley2021towards}. In this paper, we propose a data-driven approach for a more fine-grained representation of explainability and propose a user study evaluation instrument to create a metric that models explainability as a function of several sub-factors, enabling direct comparison between systems and targeted improvements.

\section{Study Design} \label{study_design}

\subsection{Questionnaire Design}

First, to compile a list of candidate aspects that may potentially contribute to the composite notion of explainability, we conducted a comprehensive structured literature review, and include the most commonly discussed aspects of explainability. We included the proceedings of ML, IR, natural language processing (NLP), and human-computer interaction (HCI) venues (i.e. ACL, CHI, ICML, NeurIPS, SIGIR) and noted papers for further review if titles included the keywords \textit{interpretability}, \textit{explainability}, or \textit{transparency}, and cross-referenced papers using \textit{connectedpapers.com} to find similar papers, resulting in 44 papers (37 of which were published within the last 7 years). We then read abstracts and conclusions for this pool to retain only those papers that examined some concrete element or aspect of explainability/interpretability, leaving us with 14 papers covering 26 unique aspects of explainability (i.e., \textit{trustworthiness}, \textit{uncertainty}, \textit{faithfulness}) (full list in Table \ref{tab:aspects}). Our final number of candidate aspects is consistent with, and perhaps more encompassing than, other survey papers such as \citet{nauta2022anecdotal}, who find 12 explainability factors from the literature. Given the flexibility of our framework, future work could easily investigate additional aspects from broader literature.

Next, these aspects were turned into a set of concrete questions (referred to as ``items'' in psychometrics) to be included in the questionnaire. We recorded responses on a 7-point Likert scale ranging from 1 (Strongly Disagree), via 4 (Neutral), to 7 (Strongly Agree). Our questionnaire was created using the following guidelines \citep{furr2011scale}: (1) items should use clear language and avoid complex words, (2) items should not be leading or presumptuous, and (3) the instrument should include both positively and negatively keyed items.

Additionally, as explainability relies on both system and explanation perception, we created items taking both into account, so our final evaluation would reflect these desiderata. To combat fatigue effects, we chose to create 2 items per aspect (one positively and one negatively worded), for a total of 52 items presented in fully randomized order, with the expectation that the discovery of latent factor representations during factor analysis would establish groupings of multiple related items. 11 doctoral and post-doctoral researchers reviewed our questionnaire for clarity and accuracy given aspect definitions. From this evaluation, we were able to identify and correct potential inconsistencies before our pilot study.

\begin{table*}
  \caption{Candidate Aspects}
  \begin{tabular}{lll}
    \toprule
    Aspect & Definition & Sources\\
    \midrule
    Simulatability & \begin{tabular}[c]{@{}l@{}}Ability to step through a system w/o a computer for a given input and produce \\the correct output\end{tabular}  & \cite{lipton2018mythos, slack2019assessing, arrieta2020explainable} \\
    Decomposability & Each part of a system’s components can be understood and explained & \cite{lipton2018mythos, arrieta2020explainable} \\
    Algorithmic Transparency & Understanding the system’s learning algorithm & \cite{lipton2018mythos, arrieta2020explainable}\\
    Causality & Ability to infer causal relationships from observational data & \cite{lipton2018mythos, arrieta2020explainable} \\
    Uncertainty & How confident the model is in its prediction & \cite{bhatt2021uncertainty, doshi2017towards, doshi2018considerations}\\
    Immediacy & \begin{tabular}[c]{@{}l@{}}When a search query is modified, how quickly outcomes of the query are displayed \\to the user\end{tabular} &  \cite{schnabel2020impact}\\
    Visibility & \begin{tabular}[c]{@{}l@{}}When a search query is slightly modified, how changes in the ranking are presented \\to the user\end{tabular} & \cite{schnabel2020impact, qvarfordt2013looking}\\
    Transferability & Ability to use system in different search contexts & \cite{lipton2018mythos, arrieta2020explainable}\\
    Model Fairness & Model makes fair/ethical decisions & \cite{lipton2018mythos,arrieta2020explainable} \\
    Understandability & Result interface presents rankings in a manner that can be understood easily by users & \cite{bibal2016interpretability}\\
    Informativeness & Provides useful information for task & \cite{lipton2018mythos, arrieta2020explainable}\\
    Global Interpretability & Knowing general factors that contribute to ranking results & \cite{doshi2017towards, doshi2018considerations, qvarfordt2013looking}\\
    Local Interpretability & Knowing the reasons for specific rankings & \cite{doshi2017towards, doshi2018considerations, qvarfordt2013looking}\\
    Counterfactuals & Ability to correctly determine how small changes to a query will affect ranking results & \cite{slack2019assessing}\\
    Efficiency & Time spent understanding the interface & \cite{bibal2016interpretability}\\
    Criticism & Knowing where and how the search engine may fail to explain certain data points & \cite{kim2016examples}\\
    Compositionality & Structure of result interface & \cite{doshi2017towards, doshi2018considerations, lage2019human}\\
    Units of Explanation & Form and number of cognitive chunks & \cite{doshi2017towards, doshi2018considerations, lage2019human}\\
    Acceptability & Accepted for use  & \cite{bibal2016interpretability}\\
    Faithfulness & How accurately the interface reflects the true reasoning process of the search engine & \cite{jacovi2020towards}\\
    Plausibility & How convincing the ranking results are to users & \cite{jacovi2020towards}\\
    Accuracy & How well the interface describes how the search engine ranked the results & \cite{jacovi2020towards}\\
    Completeness & \begin{tabular}[c]{@{}l@{}}Result interface provides accurate and complete descriptions of the search engine's \\operations\end{tabular} & \cite{gilpin2018explaining}\\
    Trustworthiness & Confidence in ranking result accuracy & \cite{lipton2018mythos, arrieta2020explainable, polley2021towards}\\
    Justifiability & System produces results that align with human expert judgements & \cite{bibal2016interpretability}\\
    Explanation Fairness & System is accessible and fair towards all people & \cite{jacovi2020towards}\\
    \bottomrule
  \end{tabular}
\label{tab:aspects}
\end{table*}

\subsection{Task Setup}

Participants performed 3 search tasks distributed across 3 topics. To motivate and guide their search, users were asked to answer a multiple-choice question for each topic. We provided users with a mock search engine based on current, non-transparent, commercial search engines that displayed a query and a list of results. We hosted our site on Netlify and displayed the task on MTurk.

Topics and questions were selected from the TREC 2004 Robust Track Dataset \citep{voorhees_overview_2005}, comprising documents from the Federal Register, Financial Times, Foreign Broadcast Information Service, and LA Times. Multiple choice answers were created by the authors to ensure they were not on the first page of results and required clicking into documents. To ensure there would be enough relevant documents to populate the results page, we randomly sampled 9 topics that contained at least 50 relevant documents. Topics were grouped as follows: \textbf{(A)} industrial espionage; income tax evasion; in vitro fertilization, \textbf{(B)} radioactive waste; behavioral genetics; drugs in the Golden Triangle, \textbf{(C)} law enforcement, dogs; non-US media bias; gasoline tax in US. For each topic, we presented 100 pre-selected documents with a 50/50 random sample of relevant and non-relevant documents in a randomized ranking order, requiring workers to interact with the search system in order to successfully complete the multiple-choice quiz.

\begin{figure}[h!]
    \centering
    \includegraphics[width=0.35\textwidth]{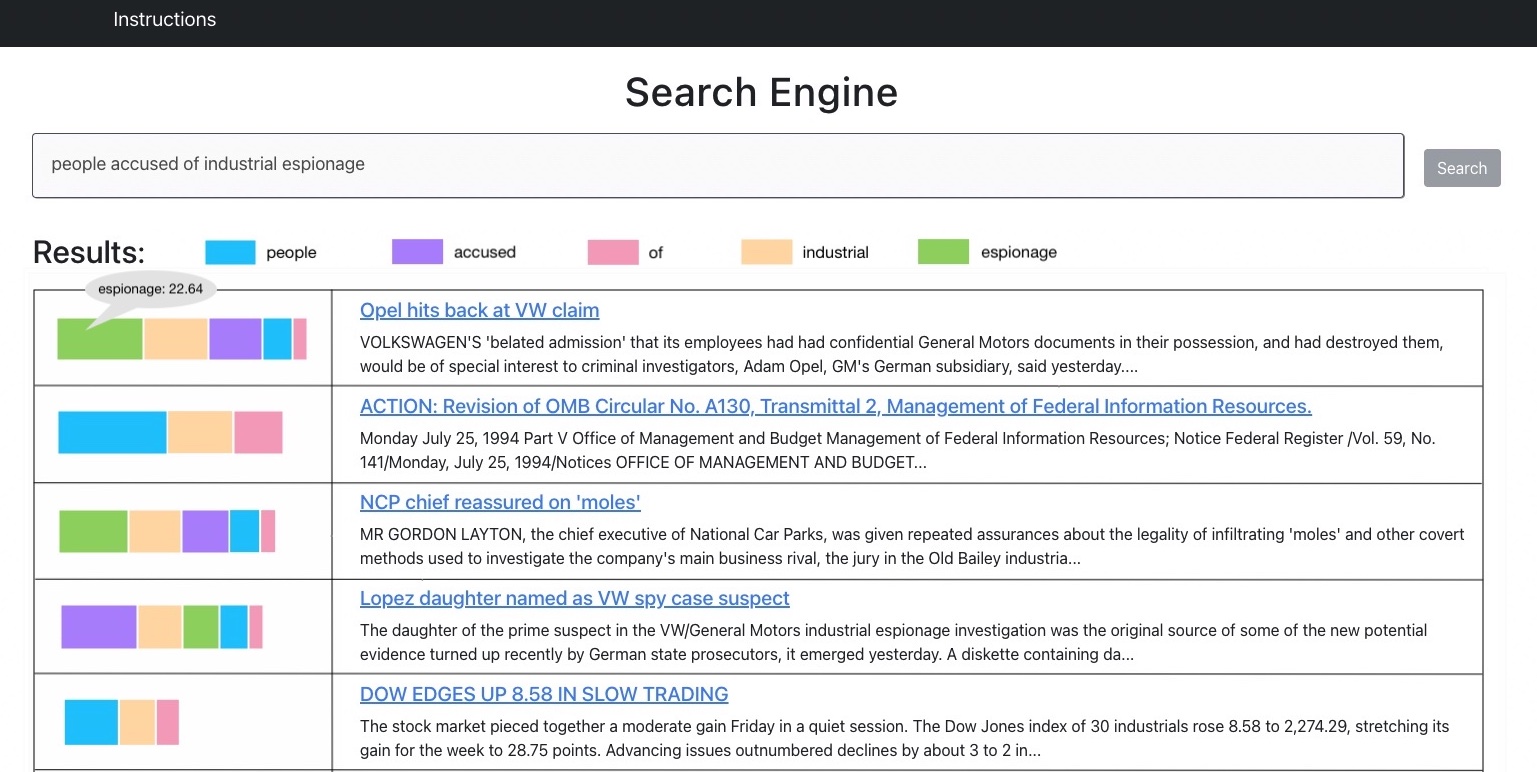}
    \caption{Search interface shown to Group A (modeled on the basis of the system presented by Ramos and Eickhoff \citep{ramos2020search}). On the left-hand side, the stacked bar graphs depict hypothetical scores of each keyword in the query for each respective search result. The larger the stacked bar graph, the more relevant that result is to the query.}
\label{fig:interface_a}
\end{figure}

Participants were randomly assigned a topic grouping. After completing the search task, we presented each group with another mock search interface based on existing transparent search systems from the literature \citep{ramos2020search, cohen2021not}, designed to test varying degrees of explainability, and participants were asked to complete our questionnaire for this second system. To avoid priming effects and other potential biases, we employed a between-subjects study design. Group \textbf{A} participants received an interface that provided visual explanation aids, with stacked bar graphs displayed next to each search result that informed users how much each query term influenced the corresponding document ranking (Figure \ref{fig:interface_a}). Group \textbf{B} participants received an interface that displayed relevance and confidence scores for each result, where confidence was modeled as a function of uncertainty (i.e., confidence and relevance percentages shown next to each result). Group \textbf{C} participants received the same non-transparent system they interacted with during the screening search task. Each condition was accompanied by brief usage instructions explaining the novel (if any) interface features.  

We included multiple quality control checks to verify worker attentiveness. Specifically, we monitored site interactions (number of clicks, documents viewed, time spent), employed a multiple-choice quiz, and provided a unique code for the worker to submit on MTurk for task completion verification. In addition to serving as a form of quality control, the multiple choice quiz was used to help guide the workers through the search task and foster a search mindset, enhancing the accuracy of the questionnaire completion.

While users’ familiarity with topics might impact their experience during the search task, the main results of this study are drawn from the questionnaire experience, which (1) was separate from the search task where the topics were presented and (2) users were asked to comment on the nature of a system, not the search task they previously performed. The questionnaire was intended to capture the extent of perceived system explainability.

\subsection{Pilot Study}

We collected 62 responses from MTurk workers over a two-month period during our pilot study. Observations during this phase influenced changes in our task design. We added pop-up confirmations to remind workers of experiment rules due to some misreading or skipping of instructions. Additionally, we implemented an early exit in the workflow and treated the search task and multiple choice quiz as a prerequisite for our survey to filter out workers who did not faithfully attempt our task. Finally, we added a uniqueness constraint to block workers from attempting our task multiple times.

Additionally, we received feedback from workers that our initial time limit (45 min) felt too rushed, leading us to increase the timer to 1 hour for our full study. However, we found that most workers spent less than the initial time limit on our task (averaging approximately 30 min). We paid workers \$9.20 for the original expected work time of 45 minutes, the equivalent of the legal minimum wage in our location. We required that workers have more than 10,000 prior approved HITs with an approval rate greater than 98\%.

\section{Data Analysis} \label{data_analysis}

We collected a total of 540 responses from our main study (Group A: 202, Group B: 134, Group C: 201)\footnote{There is a slight imbalance despite conditions being randomly assigned, but distributions are roughly preserved across groups before and after preprocessing.}. We filtered out 81 responses (15\%) during our preprocessing stage to account for workers who passed our initial quality control checks during the search task but recorded inattentive or careless responses in the subsequent survey. Following guidelines for identifying careless responses \citep{johnson2005ascertaining, behrend2011viability, meade2012identifying, goldberg1985prediction}, we analyzed response patterns and self-consistency.

Concretely, we filtered out responses that had (1) abnormally long unbroken strings (i.e., length $> 8$) of identical responses (e.g., a respondent answering a series of 18 consecutive questions with the same Likert-scale rating), (2) high overall numbers of inconsistent responses for positively and negatively keyed item pairs  (i.e., total pairs $> 4$), and (3) high amounts (i.e., total $> 5$) of responses that were more than 2 points apart for highly similar question pairs. Additionally, we filtered out items from 5 aspects (i.e., \textit{immediacy}, \textit{efficiency}, \textit{criticism}, \textit{completeness}, \textit{explanation fairness}) that produced inconsistent responses across all users. Unlike previous work that sometimes imposed even stricter criteria for preprocessing, we relaxed thresholds due to our overall task and survey lengths; we believe that workers who faithfully completed our task may not have any intentional carelessness, but instead, as they see more questions in the survey, their chance likelihood of giving a single inconsistent answer should be accounted for. Additionally, we also checked for potential ordering effects in the response data and found none (Pearson's r=-0.012 between the position at which an item was presented and its response). A total of 459 valid responses were retained after filtering. Best practices recommend a minimum sample size of 150 for EFA and 200 for CFA \citep{worthington2006scale, tabachnick2007using}. The 459 responses (Group A: 176, Group B: 110, Group C: 173) were randomly split into 2 sets: 200 responses for EFA and 259 responses for CFA.\footnote{For reproducibility purposes, we make all study code and data publicly available at \url{https://github.com/catherineschen/sse-metric}.} 

\subsection{Exploratory Factor Analysis (EFA)}

\definecolor{light-gray}{gray}{0.80}

\begin{table*}[]
    \caption{Questionnaire items and EFA factor loadings}
    \centering
    \begin{tabular}{cccl}
        \toprule
        \multicolumn{2}{c}{Factor} & \multirow{2}{*}{\begin{tabular}[c]{@{}c@{}}h\textsuperscript{2}\\\end{tabular}} & \multirow{2}{*}{Questionnaire Item} \\ 
        \cmidrule{1-2}
        1 & \multicolumn{1}{c}{2} & & \\ 
        \midrule
        \cellcolor{light-gray}{0.94} & 0.21 & 0.93 & \begin{tabular}[c]{@{}l@{}}  \textbf{15.} This system would work well in a different search task (i.e. looking up medical papers to diagnose a\\ patient).  \end{tabular} \\
        \cellcolor{light-gray}{0.84} & -0.05 & 0.71 & \textbf{37.} I would use this search engine in my everyday life. \\
        \cellcolor{light-gray}{0.83} & -0.10 & 0.70 & \textbf{21.} The results page provides me enough information to find the answers I am looking for effectively. \\
        \cellcolor{light-gray}{0.78} & -0.15 & 0.62 & \textbf{41.} The presentation of the results leads me to believe the results are ordered correctly. \\
        \cellcolor{light-gray}{0.76} & -0.20 & 0.62 & \textbf{49.} The results match my expectations and I agree with them. \\
        \cellcolor{light-gray}{0.75} & -0.21 & 0.61 & \textbf{47.} I trust that the results are ordered correctly and system will order results correctly for other queries. \\
        \cellcolor{light-gray}{0.54} & -0.34 & 0.41 & \textbf{19.} I can easily understand the contents of the results page. \\
        
        0.13 & \cellcolor{light-gray}{0.93} & 0.88 & \textbf{26.} If I change the query, I do not know how it will affect the result ordering. \\
        -0.06 & \cellcolor{light-gray}{0.89} & 0.79 & \begin{tabular}[c]{@{}l@{}} \textbf{0.} I do not understand why the results are ordered the way they are and would not be able to recreate \\ the orderings myself.\end{tabular} \\
        -0.01 & \cellcolor{light-gray}{0.87} & 0.76 & \textbf{6.} I think I need more information to understand why the given query produced the displayed results. \\
        -0.06 & \cellcolor{light-gray}{0.87} & 0.76 & \textbf{22.} I do not understand the document properties that cause some results to be ordered higher than others. \\
        0.11 & \cellcolor{light-gray}{0.87} & 0.77 & \textbf{12.} I am unable to see and understand how changes in the query affect the result ordering. \\
        -0.16 & \cellcolor{light-gray}{0.78} & 0.63 & \begin{tabular}[c]{@{}l@{}} \textbf{38.} The result interface does not help me understand the true decision making process of the search \\ engine ranker.\end{tabular}  \\
        -0.16 & \cellcolor{light-gray}{0.77} & 0.61 & \textbf{24.} I do not understand why each result is ordered in a certain place. \\
        -0.20 & \cellcolor{light-gray}{0.75} & 0.60 & \textbf{4.} I'm unable to follow how the search engine ordered the results. \\
        -0.16 & \cellcolor{light-gray}{0.74} & 0.57 & \begin{tabular}[c]{@{}l@{}} \textbf{2.} It's difficult for me to break down each of the search engine's components and understand why the \\ results are ordered the way they are.\end{tabular} \\
        -0.12 & \cellcolor{light-gray}{0.66} & 0.45 & \textbf{8.} I do not know how confident the search engine is that its displayed orderings are correct. \\
        -0.27 & \cellcolor{light-gray}{0.64} & 0.49 & \begin{tabular}[c]{@{}l@{}}  \textbf{46.} I do not trust that the results are ordered correctly and that the system will correctly order results for \\ other queries.\end{tabular}  \\
        -0.24 & \cellcolor{light-gray}{0.63} & 0.45 & \begin{tabular}[c]{@{}l@{}}  \textbf{34.} The format and amount of information provided in the result interface is not enough to help me\\ understand why the results are ordered the way they are. \end{tabular} \\
        \bottomrule
    \end{tabular}
\label{tab:efa_results}
\end{table*}

EFA is used to identify latent factors and associated questionnaire items by examining covariances in the observed data and grouping correlated items into factors \citep{tabachnick2007using}. 

To ensure the adequacy of our sample size of 200 for EFA, we followed the accepted practice of using  Barlett's Test of Sphericity and Kaiser-Meyer-Olkin (KMO) Measure of Sampling Adequacy \citep{worthington2006scale}. Barlett's Test determines if correlations between items are large enough for reduction by comparing the correlation matrix to the identity matrix. Statistically significant results indicate that the data is suitable for factor analysis as the correlation matrix is not orthogonal. However, this test is sensitive to sample size, and additional evidence is recommended to show factorability \citep{tabachnick2007using}. To supplement Barlett's Test, KMO measures the proportion of variance among variables that may be attributed to common variance. \citet{tabachnick2007using} recommend that results be greater than 0.6. Barlett's test resulted in a value of 11069.50 ($p<0.001$) and KMO resulted in a value of 0.98, indicating suitability for factor analysis.

To extract factors, we used factor analysis, which is suitable for understanding latent constructs that contribute to variance among observations and for scale development, unlike other methods such as PCA \citep{worthington2006scale, costello2005best}. Specifically, we employed principal axis factoring (PAF), a least squares estimation of the latent factor model that minimizes the sum of the ordinary least squares \citep{de2012factor, cudeck2007factor}.
    
Since human behavior is rarely independent between functions, we assumed observations to be correlated and applied the appropriate oblique (promax) rotation to improve interpretability and clarify the factor solution structure by maximizing high item loadings and minimizing low item loadings \citep{worthington2006scale, tabachnick2007using, williams2010exploratory}. Oblique rotation allows for inter-factor correlation, versus the alternative orthogonal rotation, which produces uncorrelated factors. When factors are entirely uncorrelated, both methods yield similar results \citep{costello2005best}.
    
The most popular methods to determine the number of factors to preserve include the Scree test \citep{cattell1966scree}, Kaiser's criterion \citep{kaiser1960application}, or parallel analysis \citep{horn1965rationale}. However, there is no clear consensus in the literature on which method is most reliable. Kaiser's criterion suggests retaining those factors whose eigenvalues are greater than 1.0, which suggested we should retain 2 factors. Scree plot examination can often be unclear since it calls for visual inspection to determine a ``leveling off'' point in the graph, and thus, its subjective nature can make the test unreliable. Parallel analysis involves calculating eigenvalues from a randomly generated dataset and comparing the values to the observed matrix, which indicated we should keep one factor. Overall, since no consensus was drawn from these methods, we tested both one and two-factor solutions (i.e., first-order and hierarchical two-factor models) in Section \ref{cfa}.

We conducted an additional round of EFA fixing the number of factors to two and discarded items with weak factor loadings (<0.4), large cross-loading differences (>0.15), large absolute multiple factor loadings (>0.4), or weak communality $ h^2 $ (<0.4), as suggested by \citep{comrey2013first, worthington2006scale, tabachnick2007using}. It is also important to note that at this stage, it is suggested to approximate a \textit{simple structure} \citep{thurstone1947multiple}, meaning that factor groupings should seek to have intuitive meaning and items should only load on a single factor. To achieve this, researchers have suggested retaining at least three items per factor, deleting misfitting items, or even repeating the study with additional items that are hypothesized to contribute to a specific factor \citep{tabachnick2007using, worthington2006scale}. Criteria for factor extraction and retention should not be interpreted as a strict rule, but instead, interpretability and other practical considerations should be taken into high account \citep{auerswald2019determine, worthington2006scale}.

We conducted a final round of EFA to confirm the factor solution stability after item deletion and analyzed the groupings for interpretability. Table \ref{tab:efa_results} shows the final proposed item groupings and resulting factor loadings. Out of our original pool of 52, we retain twelve items in Factor Group 1 and seven items in Factor Group 2.

\subsection{Confirmatory Factor Analysis (CFA)} \label{cfa}

\begin{table*}
    \caption{Global fit statistics for CFA}
    \begin{tabular}{lccccccc}
        \toprule
        Model & $\chi^2 \downarrow$ & $df \downarrow$ & $\chi^2$ / $df \downarrow$ & CFI $\uparrow$ & NNFI $\uparrow$ & RMSEA $\downarrow$ & SRMR $\downarrow$ \\
        \midrule
        Null model & 5792.80 & 171 & 33.88 &-- & -- & -- & -- \\
        First-order model & 465.08 & 152 & 3.06 & 0.937 & 0.944 & 0.089 & 0.026 \\
        Hierarchical two-factor model & \textbf{327.55} & \textbf{150} & \textbf{2.18} & \textbf{0.968} & \textbf{0.964} & \textbf{0.068} & \textbf{0.021} \\
        \bottomrule
    \end{tabular}
\label{tab:cfa_results}
\end{table*}

While EFA determines a candidate model structure, CFA confirms the EFA-derived model fit. Goodness of fit is assessed by examining the alignment between the model-estimated and observed covariance matrices \citep{ullman2012structural, brown2012confirmatory} on a held-out set of data. SEM is commonly used to confirm the fit of potential model structures. To approximate the covariance matrix, we followed standard practice of using maximum likelihood estimation, which maximizes the likelihood that the model estimated parameters fit the observed data \citep{ullman2012structural, de2012factor}.

\begin{figure}[h]
    \centering
    \includegraphics[width=0.35\textwidth]{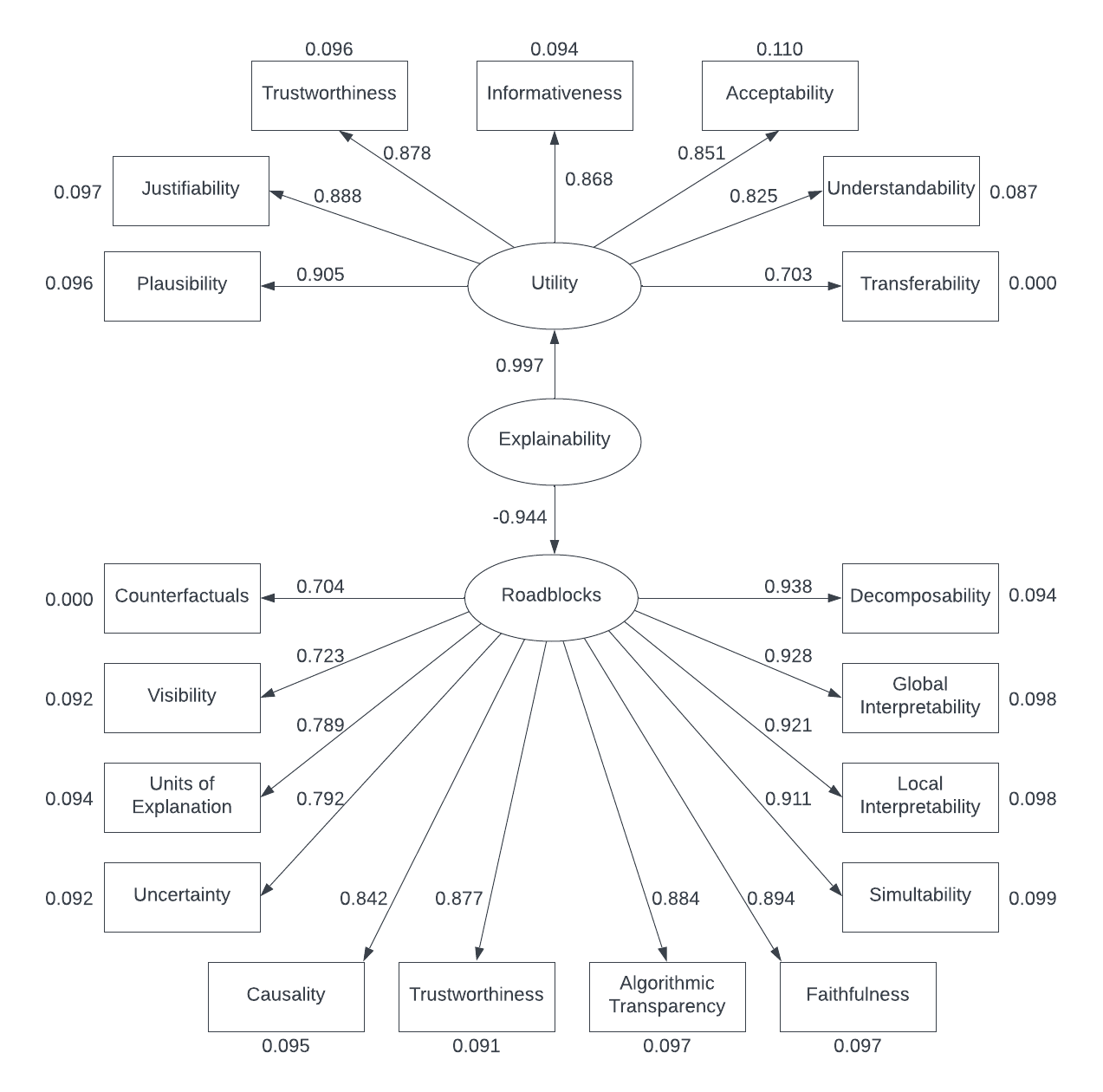}
    \caption{\textit{Path diagram} for proposed structural equation model for modeling explainability.}
\label{fig:hmodel}
\end{figure}

We tested our EFA-derived hierarchical two-factor model on a held-out data set of 259 responses, which exceeded the recommended minimum size of 200 \citep{worthington2006scale, maccallum1999sample}. We also compared our proposed model to a null model (all items assumed independent with covariances fixed to 0), and a first-order factor model (all items loaded onto a single latent factor). Figure \ref{fig:hmodel} shows a visual representation of our hierarchical factor model using a \textit{path diagram}, which depicts the relationships between latent factors and items in SEM. Factors appear in circles or ovals, and items in squares or rectangles. Arrows connect entities, with single-headed arrows indicating direct relationships (with one variable loading on the other), and double-headed arrows indicating relationships without direction (with covariance represented by edge weight). 

Common research practice in SEM is to use a chi-square goodness of fit test, however, due to its sensitivity to sample size, it is suggested to supplement this test with alternative fit indices \citep{hu1999cutoff, bentler1980significance, kline2005psychological}. Some researchers report a relative chi-square statistic ($\chi^2$ / $df$) to minimize the impact of sample size, but there is no strong consensus on acceptable values, ranging from 5.0 \citep{wheaton1977assessing} to 2.0 \citep{tabachnick2007using}. We report the results of both chi-square statistics for completeness, but followed standard practice of assessing model fit by examining two additional categories of fit indices: absolute fit to measure how well our model fit the observed data and incremental fit to measure our proposed model against a baseline model \citep{hooper2008structural, kline2005psychological, hu1999cutoff}. Table \ref{tab:cfa_results} shows that our hypothesized hierarchical two-factor model achieved a better fit over the null and first-order models, with all fit statistics well within the acceptable ranges. For incremental fit indices, we report the Comparative Fit Index (CFI) and the Non-Normed fit index (NNFI), also known as the Tucker Lewis Index (TLI), with values at 0.968 and 0.964, respectively, both above the standard acceptable value of 0.95 \citep{hu1999cutoff}. For absolute fit indices, we supplemented the chi-square test with the root mean square error of approximation (RMSEA) and the standardized root mean square residual (SRMR), with values at 0.068 and 0.021, respectively, below the acceptable levels of 0.07 and 0.08 \citep{steiger2007understanding, hu1999cutoff}.

\subsection{Explainability Factor Composition}
Here, we discuss which aspects respondents found most important for explainability. The results of CFA confirm that items organize into two distinct factor groups (Figure \ref{fig:hmodel}). Examining the individual items reveals that constructs are separated into positive and negative traits. In Factor 1, we retain seven items referring to positive attributes and in Factor 2, we retain twelve items referring to negative attributes. We name these factor groups \textbf{(1)} \textit{utility} and \textbf{(2)} \textit{roadblocks}. The first corresponds to the system's \textit{utility}, falling in line with existing evaluation strategies where system explainability is tested via its usefulness within some context or application \citep{ribeiro2016should, lundberg2017unified}. The second compiles a range of critical \textit{roadblocks} that can be thought of as properties a system might lack in order to be fully explainable. In Figure \ref{fig:hmodel}, we also show the original aspect labels associated with each item, derived from our literature review during the questionnaire development phase. Intuitively, item-factor loadings represent how significant an item is to the overall factor. For example, Item 41 (\textit{plausibility}) is a strong indicator of Factor 1 (\textit{utility}). In other words, we can interpret this as the higher the \textit{plausibility} score, the more useful its explanations are. Conversely, Item 2 (\textit{decomposability}) is a strong indicator of Factor 2 (\textit{roadblocks}). Since Factor 2 has a negative loading, higher responses on Item 2 indicate a strong \textit{lack of decomposability} and, thereby lower explanation capability.

\section{Search System Explainability (SSE)} \label{metric}

We extend this factor analysis to introduce Search System Explainability (SSE), a metric for evaluating search system explainability. In Equation \ref{eq: score}, $F$ is the set of all factors $f$ and $I_f$ is the set of survey items $i$ that correspond to factor $f$. The user's response $r_i$ to item $i$ is weighted by the corresponding loading coefficient $w_i$  and the accumulated response scores for each factor are weighted by the corresponding loading coefficient $w_f$ determined by CFA. To make SSE score values more interpretable, min-max normalization with the smallest and largest theoretically possible SSE scores is applied to normalize overall values between 0 and 1.

    \begin{equation} \label{eq: score}
        SSE = MinMax Norm \bigg( \sum_{f\in{F}} w_f \sum_{i\in{I_f}} w_i r_i \bigg)
    \end{equation}

\subsection{Evaluation Setup}

In this section, we aim to test the hypothesis that an explainable system will attain a higher SSE score than a non-explainable system.
We employed a similar task setup to our previous crowdsourcing task, with a few differences:
\begin{itemize}
    \item Users were asked to evaluate 1 of 2 systems: (A) \texttt{BASELINE}: a minimalist search engine resembling existing commercial search systems with no explainable capabilities and (B) \texttt{BARS}: a search system with visual explanations representing query term importance resembling the interface shown to Group A in Section \ref{study_design}.
    \item Users were asked to answer an open-ended question on one of ten possible news topics\footnote{List of topics: compost pile, Antarctica exploration, heroic acts, Northern Ireland industry, new hydroelectric projects, health and computer terminals, quilts and income, wildlife extinction, illegal technology transfer, Salvation Army benefits} using a different document set (still from TREC Robust 2004).
    \item Both search systems were fully interactive instead of static and used the same BM25 retrieval model to rank documents.
\end{itemize}

After completing the search task, participants were again asked to fill out the evaluation questionnaire. We incorporated two attention check items (``\textit{I swim across the Atlantic Ocean to get to work every day.}'' and ``\textit{I think search engines are cool. Regardless of your opinion, select `Strongly Agree' as your response below. This is an attention check.}'') at random points for quality control.

Following the evaluation questionnaire, we administered the NASA Task Load Index (NASA-TLX) \cite{hart2006nasa} to measure the perceived workload of the evaluation questionnaire. The NASA-TLX is a 6-item questionnaire measuring perceived workload on 6 dimensions (\textit{Mental Demand, Physical Demand, Temporal Demand, Performance, Effort, Frustration}) with responses recorded on a scale of 0-20, where lower scores are more desirable and indicate a low workload (except for Performance, where lower scores indicate high success). Since the original scoring method involves a two-step weighting scheme through numerous pairwise comparisons, we followed the common practice of analyzing raw TLX (RTLX) responses to reduce the amount of time needed to administer the survey \cite{hart2006nasa}. 

Prior to full-scale data collection, we conducted a think-aloud study with 6 graduate students to identify and address flaws in the study design and system. We used the concurrent think-aloud method due to its effectiveness in usability testing and time efficiency \cite{olmsted2010think}. Observations from this period led to several changes. Specifically, we made slight modifications to the NASA-TLX assessment: (1) we replaced the term `task' with `survey' to align responses with the evaluation questionnaire rather than the search task and (2) we inverted the scale on the \textit{Performance} dimension to have higher responses equate to `Perfect' rather than `Failure' and align with a more intuitive understanding of performance. Additionally, we added more detailed descriptions to certain questions to accurately anchor neutral responses of low effort to 0 instead of 10, which would indicate a medium effort.

We recruited 100 participants aged 18-65 for our study on Prolific\footnote{IRB approval was judged unnecessary by an IRB member due to the classification of our study as a system evaluation.} and paid them £5.25 for 30 minutes of work (estimated from a 15-participant pilot study), equivalent to the minimum wage in our location. 
On average, users completed the task more quickly than our initial estimate, with the median time being approximately 22 minutes. 15 countries were represented in our sample, with participants from Australia (1), Africa (4), Europe (81), North America (12), South America (1), and 1 undisclosed location.

\subsection{Results}

\begin{figure}[h!]
    \centering
    \includegraphics[width=0.30\textwidth]{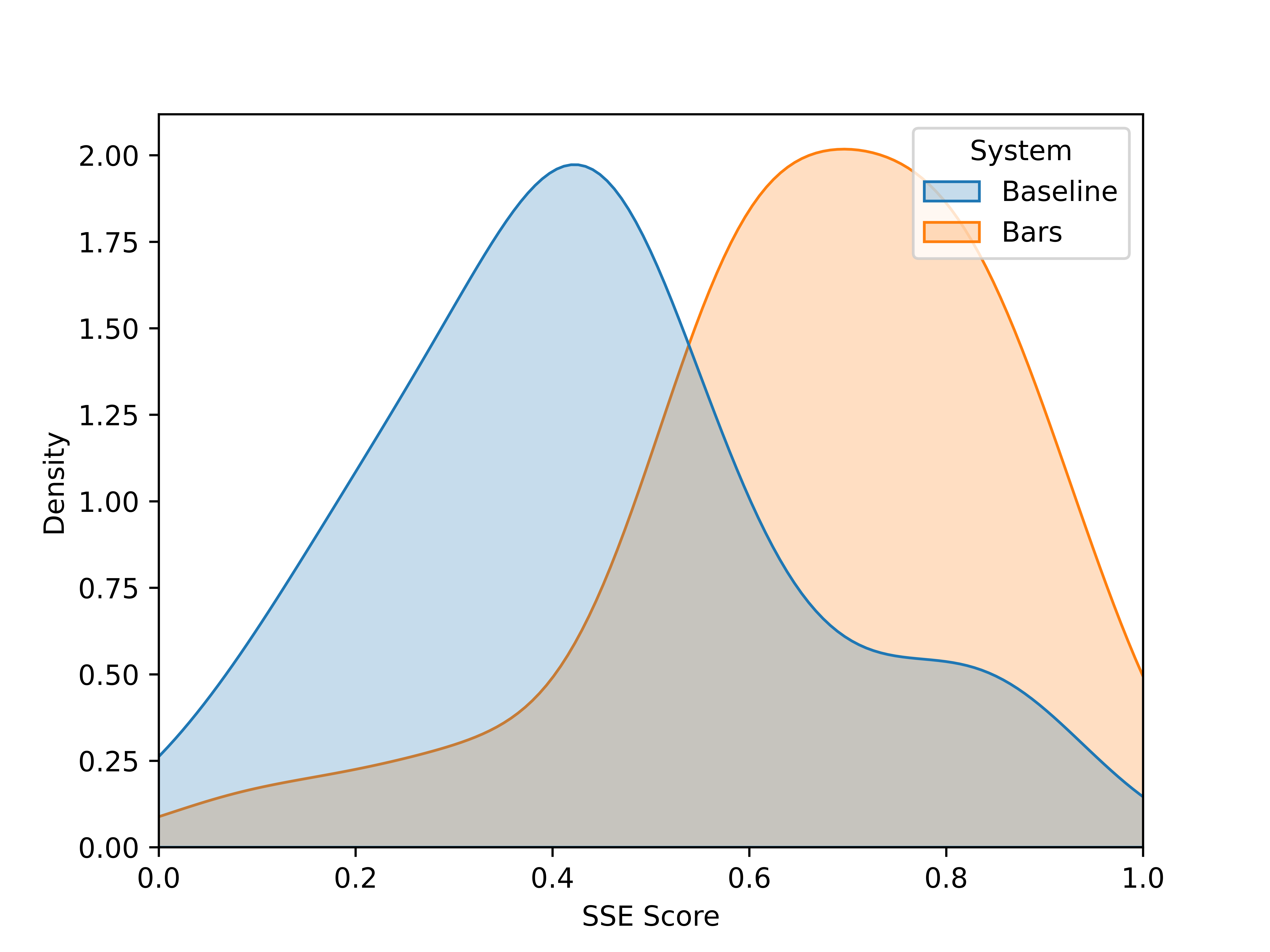}
    \caption{Distribution of SSE scores. Results from a Wilcoxon signed rank test (T = 98.0, p < 0.001) indicate there is a statistically significant difference in scores between systems.}
    \label{fig:sse_scores}
\end{figure}

\noindent\textbf{System Explainability.}
Congruent with our initial hypothesis, we find that users assigned higher explainability scores to the \texttt{BARS} system in comparison to the non-transparent \texttt{BASELINE} system (Figure \ref{fig:sse_scores}). The \texttt{BARS} system exhibited a mean SSE score (M=0.67) greater than that of the \texttt{BASELINE} system (M=0.44). The results from a Wilcoxon signed rank test on a random sample of the maximum number of samples in the minority class (n=44) reveal a statistically significant difference (T=98.0, p<0.001), confirming that SSE is able to accurately discern between explainable and non-explainable systems.

\noindent\textbf{Search Efficiency.} 
Despite users in the \texttt{BARS} group spending a slightly longer duration on the overall search task (11.97 min) compared to the \texttt{BASELINE} group (11.41 min), an analysis of search time per query and the number of viewed documents per query indicates that the explainable \texttt{BARS} system exhibited higher per-query efficiency than the non-transparent system (Table \ref{tab:efficiency}). Users issued more queries per task in the explainable system (3.45 vs 2.54). This may be attributed to query modifications aimed at targeting specific keywords to leverage query-term matching for calculating BM25 scores. These results suggest that the \texttt{BARS} system allowed users to scan search engine result pages (SERP) more quickly. This observation aligns with previous research by \citet{ramos2020search}, which demonstrated the positive impact of explainable systems on search efficiency.

\begin{table}[h!]
\caption{Task engagement statistics (M $\pm$ SD). Although participants in the \texttt{BARS} group spent more time overall compared to the \texttt{BASELINE} group, they were more efficient per SERP.}
\begin{tabular}{@{}lrr@{}}
\toprule
                           & \texttt{BASELINE}   & \texttt{BARS}         \\ 
                           & n=56               & n=44 \\
\midrule
Task time (in min)         & 11.41 $\pm$ 8.81  & 11.97 $\pm$ 6.42 \\
\# of queries              & 2.54  $\pm$ 1.95  & 3.45  $\pm$ 2.90 \\
\# of documents viewed     & 6.30  $\pm$ 4.82  & 6.89  $\pm$ 5.56 \\
Time per query             & 6.79  $\pm$ 8.22  & 5.36  $\pm$ 4.50 \\
Documents viewed per query & 3.84  $\pm$ 4.65  & 2.91  $\pm$ 2.85 \\ 
\bottomrule
\end{tabular}
\label{tab:efficiency}
\end{table}

\begin{figure*}[h!]
    \centering
    \includegraphics[width=0.65\textwidth]{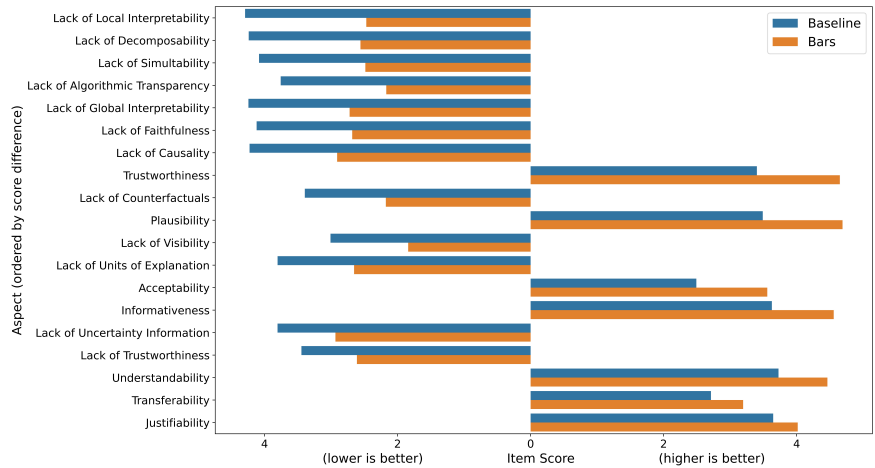}
    \caption{Loadings for individual questionnaire items. Items are labeled by their original aspect as determined and ordered by score (response multiplied by item loading) difference between the two systems. Aspects with bars increasing toward the right indicate the practical usefulness of each system, while aspects with bars increasing toward the left represent areas requiring improvement to achieve full explainability.}
    \label{fig:aspect-scores}
\end{figure*}

\section{Discussion} \label{discussion}

\subsection{Dimensions of Search Explainability}

Overall, our results support the multidimensional nature of explainability posited by recent literature \citep{doshi2018considerations, lipton2018mythos, nauta2022anecdotal}. We contribute empirical evidence that these factors group between positive and negative facets, representing the \textit{utility} and \textit{roadblocks to explainability} of search systems. While this strong split among positive and negative factors is not seen in other psychometric multidimensional modeling work in IR, such as work by \citet{zhang2014multidimensional} on multidimensional relevance, researchers posit that modeling negative aspects of user experience and reasons for non-use can be valuable in future system design \citep{noman2019techies, baumer2015importance, kim2022understanding, greisdorf2003relevance}.

In this study, we show that our proposed metric, SSE, can inform targeted system improvements and comparisons with other systems. In particular, the aspect scores offer a comprehensive evaluation of individual questionnaire items and specific areas where one system outperforms another. In Figure \ref{fig:aspect-scores}, we present a breakdown of the aspect scores in our study and show that the \texttt{BARS} system outperformed the \texttt{BASELINE} across all areas, thus indicating higher readiness for deployment and a reduced need for extensive feature modifications to achieve full explainable capability. Notably, items are grouped by their representation of attributes relating to utility (right) and areas for improvement (left). Additionally, the aspects positioned closer to the top of the graph highlight the most significant disparities in average item scores between the two systems. For instance, the widest margin between the two systems is on \textit{lack of local interpretability}, indicating that the \texttt{BARS} system has more \textit{local interpretability} than the \texttt{BASELINE} system, but still has some room for improvement. Overall, our work is a step towards deploying explainable interfaces that cause minimal disruption to the user experience by introducing our model and questionnaire as an evaluation framework that will enable targeted system improvements and comparisons with other systems.

\subsection{Implications for Evaluation}

Though some researchers criticize the quality of crowdsourced data due to poor compensation and suggest automated evaluation \citep{hara2018data, herman2017promise}, choosing a proxy evaluation method can be challenging \citep{doshi2017towards} and results do not capture the true verdict of end users who ultimately will be using these applications. Given that explainability aims to provide insight into a model's decision process in \textit{human-understandable} terms, assessing these systems with humans will provide more impactful evaluations. However, it is important to keep in mind that such evaluation methods may be hindered if they impose excessive workload burdens on crowd workers.

\begin{table*}[h!]
\caption{NASA-RTLX responses (lower is better) for perceived evaluation questionnaire workload indicate that the \texttt{BARS} system caused less \textit{Mental} and \textit{Physical Demand}, higher perceived \textit{Performance}, and less perceived \textit{Effort} and \textit{Frustration} than the non-transparent \texttt{BASELINE} system.}
\begin{tabular}{lrrrrrr}
\toprule
               & Mental Demand & Physical Demand & Temporal Demand & Performance & Effort & Frustration \\ \midrule
Overall        & 7.53          & 2.26            & 3.44            & 4.12       & 10.01  & 3.28        \\ \midrule
\texttt{BASELINE}       & 7.79          & 2.89            & 3.02            & 4.02       & 10.18  & 4.00        \\
\texttt{BARS}           & 7.20          & 1.45            & 3.98            & 4.24       & 9.80   & 2.36        \\ \midrule
\end{tabular}
\label{tab:tlx-results}
\end{table*}

We find that our evaluation questionnaire imposes a low-to-medium additional workload across 6 workload dimensions (Table \ref{tab:tlx-results}). Participants evaluating the \texttt{BARS} system reported a slightly heightened sense of urgency (i.e., \textit{Temporal Demand}), potentially due to the nature of the search task rather than the survey itself, as users in this group took slightly longer to complete the initial search task compared to the \texttt{BASELINE} group (Table \ref{tab:efficiency}). Overall, participants from both groups considered the evaluation instrument moderately demanding in terms of \textit{Mental Demand}, moderate in \textit{Effort}, low in \textit{Physical Demand}, \textit{Temporal Demand}, and \textit{Frustration}, and high success in \textit{Performance}.

On average, non-native English speakers demonstrated higher perceived levels of \textit{Mental Demand}, \textit{Physical Demand}, and \textit{Frustration} compared to native English speakers. However, their perceived \textit{Performance} was similar to that of native English speakers while reporting less \textit{Temporal Demand} and \textit{Effort}. Nevertheless, these findings are encouraging since the overall task load for all participants ranged from low to medium, suggesting the suitability of the explainability evaluation questionnaire for both native and non-native English speakers.

\subsection{Limitations} 

Although our literature review produced 26 potential facets contributing to the overall notion of explainability, we find that users are only concerned with 19. While we cannot draw conclusions from the 5 factors that were filtered out during preprocessing due to highly inconsistent responses across all users, users discount 3 aspects entirely: \textit{compositionality} \citep{doshi2017towards, doshi2018considerations, lage2019human}, \textit{model fairness} \citep{lipton2018mythos,arrieta2020explainable}, and \textit{accuracy} \citep{jacovi2020towards}. We note that our findings only apply to Web search systems; conducting this study in other search domains (i.e., clinical abstract search) or other ML tasks (i.e., image classification) may find these aspects to be important to explainability. Explainability is highly domain-specific and can change depending on the intended user and information need. For example, a clinician using a diagnostic decision support system may require more detailed information and prefer different system attributes than a journalist fact-checking sources, or a consumer comparing products. Thus, we make the distinction that the dimensions we find pertinent to explainability in this paper are limited to information-seeking needs satisfied via Web search systems by everyday laypersons, and may not hold true for all conceivable domains. However, the methodology outlined in this work can be used to create multidimensional models of explainability for other domains and tasks.

Additionally, while there is no ideal label for these overarching concepts as they are composites of multiple aspects, we attribute names that we feel encompass the nature of the items in the respective groups for a clearer, more targeted discussion of properties. 

\subsection{Ethical Considerations}

Our study employs stringent but necessary controls in order to collect faithful responses and mitigate data quality concerns. While it is possible such controls could introduce a bias towards demographics that are more willing to complete a longer task, \citet{difallah2018demographics} analyzed worker propensity with demographic correlation and found that ``most demographic variables are not affected by [such] selection biases'' (with the exception that Indian workers may be overrepresented in the pool). MTurk does not release demographic information about workers nor did we set such selection criteria, choosing to follow common practice of selecting workers based on fidelity (e.g., HIT approval rates, \# of completed HITs) to improve the chance of receiving high-quality data \citep{peer2014reputation}. However, we note that HIT batches were released in the AM (ET) and most submissions were completed by EOD or very early AM the following day, possibly resulting in more responses from time zones overlapping with this posting time (i.e., Americas/Europe). 

We also note that while we compensated our crowdsourced workers based on our location's legal minimum wage, this rate may not align with minimum wages in their respective locations. However, since demographic information was not collected or available, we followed the common practice of setting a standard compensation rate, and throughout our study, we received feedback from workers showing promising engagement and understanding of our task.

Finally, it is important to acknowledge the potential risks of modeling error in other domains and among aspects. For example, the cost of error in the biomedical domain may be much higher than the cost of error in the everyday search use case we examine in this work. Additionally, getting aspects such as \textit{trustworthiness} or \textit{uncertainty} is potentially riskier than getting \textit{visibility} wrong.

\section{Conclusion} \label{conclusion}

In this paper, we establish a user-centric definition and evaluation metric of search system explainability (SSE) grounded in recent literature. Based on a large-scale crowdsourced user study and factor analysis, we show that SSE can be used to identify \textit{utility} and critical \textit{roadblock} factors in explainable and non-transparent search systems. Future work may investigate the extent to which SSE can capture granular nuances between multiple explainable systems. Additionally, a task load analysis reveals that both native and non-native English speakers found the evaluation instrument to have a low-to-medium workload. Overall, our study offers valuable insights for future researchers in the field of XIR, providing guidance on utilizing this evaluation instrument to conduct their own assessments.

\begin{acks}
We thank Charlie Lovering, Jack Merullo, Nihal Nayak, and William Rudman for the discussion and comments on this work. Additionally, we appreciate the feedback and suggestions from Shaun Wallace and Kelsey Urgo regarding the crowdsourcing setup.
\end{acks}

\bibliographystyle{ACM-Reference-Format}
\balance
\bibliography{refs}


\begin{thebibliography}{64}


\ifx \showCODEN    \undefined \def \showCODEN     #1{\unskip}     \fi
\ifx \showDOI      \undefined \def \showDOI       #1{#1}\fi
\ifx \showISBNx    \undefined \def \showISBNx     #1{\unskip}     \fi
\ifx \showISBNxiii \undefined \def \showISBNxiii  #1{\unskip}     \fi
\ifx \showISSN     \undefined \def \showISSN      #1{\unskip}     \fi
\ifx \showLCCN     \undefined \def \showLCCN      #1{\unskip}     \fi
\ifx \shownote     \undefined \def \shownote      #1{#1}          \fi
\ifx \showarticletitle \undefined \def \showarticletitle #1{#1}   \fi
\ifx \showURL      \undefined \def \showURL       {\relax}        \fi
\providecommand\bibfield[2]{#2}
\providecommand\bibinfo[2]{#2}
\providecommand\natexlab[1]{#1}
\providecommand\showeprint[2][]{arXiv:#2}

\bibitem[Adadi and Berrada(2018)]%
        {adadi2018peeking}
\bibfield{author}{\bibinfo{person}{Amina Adadi} {and} \bibinfo{person}{Mohammed Berrada}.} \bibinfo{year}{2018}\natexlab{}.
\newblock \showarticletitle{Peeking inside the black-box: a survey on explainable artificial intelligence (XAI)}.
\newblock \bibinfo{journal}{\emph{IEEE access}}  \bibinfo{volume}{6} (\bibinfo{year}{2018}), \bibinfo{pages}{52138--52160}.
\newblock


\bibitem[Arrieta et~al\mbox{.}(2020)]%
        {arrieta2020explainable}
\bibfield{author}{\bibinfo{person}{Alejandro~Barredo Arrieta}, \bibinfo{person}{Natalia D{\'\i}az-Rodr{\'\i}guez}, \bibinfo{person}{Javier Del~Ser}, \bibinfo{person}{Adrien Bennetot}, \bibinfo{person}{Siham Tabik}, \bibinfo{person}{Alberto Barbado}, \bibinfo{person}{Salvador Garc{\'\i}a}, \bibinfo{person}{Sergio Gil-L{\'o}pez}, \bibinfo{person}{Daniel Molina}, \bibinfo{person}{Richard Benjamins}, {et~al\mbox{.}}} \bibinfo{year}{2020}\natexlab{}.
\newblock \showarticletitle{Explainable Artificial Intelligence (XAI): Concepts, taxonomies, opportunities and challenges toward responsible AI}.
\newblock \bibinfo{journal}{\emph{Information Fusion}}  \bibinfo{volume}{58} (\bibinfo{year}{2020}), \bibinfo{pages}{82--115}.
\newblock


\bibitem[Auerswald and Moshagen(2019)]%
        {auerswald2019determine}
\bibfield{author}{\bibinfo{person}{Max Auerswald} {and} \bibinfo{person}{Morten Moshagen}.} \bibinfo{year}{2019}\natexlab{}.
\newblock \showarticletitle{How to determine the number of factors to retain in exploratory factor analysis: A comparison of extraction methods under realistic conditions.}
\newblock \bibinfo{journal}{\emph{Psychological methods}} \bibinfo{volume}{24}, \bibinfo{number}{4} (\bibinfo{year}{2019}), \bibinfo{pages}{468}.
\newblock


\bibitem[Baumer et~al\mbox{.}(2015)]%
        {baumer2015importance}
\bibfield{author}{\bibinfo{person}{Eric~PS Baumer}, \bibinfo{person}{Jenna Burrell}, \bibinfo{person}{Morgan~G Ames}, \bibinfo{person}{Jed~R Brubaker}, {and} \bibinfo{person}{Paul Dourish}.} \bibinfo{year}{2015}\natexlab{}.
\newblock \showarticletitle{On the importance and implications of studying technology non-use}.
\newblock \bibinfo{journal}{\emph{interactions}} \bibinfo{volume}{22}, \bibinfo{number}{2} (\bibinfo{year}{2015}), \bibinfo{pages}{52--56}.
\newblock


\bibitem[Behrend et~al\mbox{.}(2011)]%
        {behrend2011viability}
\bibfield{author}{\bibinfo{person}{Tara~S Behrend}, \bibinfo{person}{David~J Sharek}, \bibinfo{person}{Adam~W Meade}, {and} \bibinfo{person}{Eric~N Wiebe}.} \bibinfo{year}{2011}\natexlab{}.
\newblock \showarticletitle{The viability of crowdsourcing for survey research}.
\newblock \bibinfo{journal}{\emph{Behavior research methods}} \bibinfo{volume}{43}, \bibinfo{number}{3} (\bibinfo{year}{2011}), \bibinfo{pages}{800--813}.
\newblock


\bibitem[Bentler and Bonett(1980)]%
        {bentler1980significance}
\bibfield{author}{\bibinfo{person}{Peter~M Bentler} {and} \bibinfo{person}{Douglas~G Bonett}.} \bibinfo{year}{1980}\natexlab{}.
\newblock \showarticletitle{Significance tests and goodness of fit in the analysis of covariance structures.}
\newblock \bibinfo{journal}{\emph{Psychological bulletin}} \bibinfo{volume}{88}, \bibinfo{number}{3} (\bibinfo{year}{1980}), \bibinfo{pages}{588}.
\newblock


\bibitem[Bhatt et~al\mbox{.}(2021)]%
        {bhatt2021uncertainty}
\bibfield{author}{\bibinfo{person}{Umang Bhatt}, \bibinfo{person}{Javier Antor{\'a}n}, \bibinfo{person}{Yunfeng Zhang}, \bibinfo{person}{Q~Vera Liao}, \bibinfo{person}{Prasanna Sattigeri}, \bibinfo{person}{Riccardo Fogliato}, \bibinfo{person}{Gabrielle Melan{\c{c}}on}, \bibinfo{person}{Ranganath Krishnan}, \bibinfo{person}{Jason Stanley}, \bibinfo{person}{Omesh Tickoo}, {et~al\mbox{.}}} \bibinfo{year}{2021}\natexlab{}.
\newblock \showarticletitle{Uncertainty as a form of transparency: Measuring, communicating, and using uncertainty}. In \bibinfo{booktitle}{\emph{Proceedings of the 2021 AAAI/ACM Conference on AI, Ethics, and Society}}. \bibinfo{pages}{401--413}.
\newblock


\bibitem[Bibal and Fr{\'e}nay(2016)]%
        {bibal2016interpretability}
\bibfield{author}{\bibinfo{person}{Adrien Bibal} {and} \bibinfo{person}{Beno{\^\i}t Fr{\'e}nay}.} \bibinfo{year}{2016}\natexlab{}.
\newblock \showarticletitle{Interpretability of machine learning models and representations: an introduction.}. In \bibinfo{booktitle}{\emph{ESANN}}.
\newblock


\bibitem[Brown and Moore(2012)]%
        {brown2012confirmatory}
\bibfield{author}{\bibinfo{person}{Timothy~A Brown} {and} \bibinfo{person}{Michael~T Moore}.} \bibinfo{year}{2012}\natexlab{}.
\newblock \showarticletitle{Confirmatory factor analysis}.
\newblock \bibinfo{journal}{\emph{Handbook of structural equation modeling}}  \bibinfo{volume}{361} (\bibinfo{year}{2012}), \bibinfo{pages}{379}.
\newblock


\bibitem[Cattell(1966)]%
        {cattell1966scree}
\bibfield{author}{\bibinfo{person}{Raymond~B Cattell}.} \bibinfo{year}{1966}\natexlab{}.
\newblock \showarticletitle{The scree test for the number of factors}.
\newblock \bibinfo{journal}{\emph{Multivariate behavioral research}} \bibinfo{volume}{1}, \bibinfo{number}{2} (\bibinfo{year}{1966}), \bibinfo{pages}{245--276}.
\newblock


\bibitem[Cohen et~al\mbox{.}(2021)]%
        {cohen2021not}
\bibfield{author}{\bibinfo{person}{Daniel Cohen}, \bibinfo{person}{Bhaskar Mitra}, \bibinfo{person}{Oleg Lesota}, \bibinfo{person}{Navid Rekabsaz}, {and} \bibinfo{person}{Carsten Eickhoff}.} \bibinfo{year}{2021}\natexlab{}.
\newblock \showarticletitle{Not all relevance scores are equal: Efficient uncertainty and calibration modeling for deep retrieval models}. In \bibinfo{booktitle}{\emph{Proceedings of the 44th International ACM SIGIR Conference on Research and Development in Information Retrieval}}. \bibinfo{pages}{654--664}.
\newblock


\bibitem[Comrey and Lee(2013)]%
        {comrey2013first}
\bibfield{author}{\bibinfo{person}{Andrew~L Comrey} {and} \bibinfo{person}{Howard~B Lee}.} \bibinfo{year}{2013}\natexlab{}.
\newblock \bibinfo{booktitle}{\emph{A first course in factor analysis}}.
\newblock \bibinfo{publisher}{Psychology press}.
\newblock


\bibitem[Costello and Osborne(2005)]%
        {costello2005best}
\bibfield{author}{\bibinfo{person}{Anna~B Costello} {and} \bibinfo{person}{Jason Osborne}.} \bibinfo{year}{2005}\natexlab{}.
\newblock \showarticletitle{Best practices in exploratory factor analysis: Four recommendations for getting the most from your analysis}.
\newblock \bibinfo{journal}{\emph{Practical assessment, research, and evaluation}} \bibinfo{volume}{10}, \bibinfo{number}{1} (\bibinfo{year}{2005}), \bibinfo{pages}{7}.
\newblock


\bibitem[Cudeck and MacCallum(2007)]%
        {cudeck2007factor}
\bibfield{author}{\bibinfo{person}{Robert Cudeck} {and} \bibinfo{person}{Robert~C MacCallum}.} \bibinfo{year}{2007}\natexlab{}.
\newblock \bibinfo{booktitle}{\emph{Factor analysis at 100: Historical developments and future directions}}.
\newblock \bibinfo{publisher}{Routledge}.
\newblock


\bibitem[Das and Rad(2020)]%
        {das2020opportunities}
\bibfield{author}{\bibinfo{person}{Arun Das} {and} \bibinfo{person}{Paul Rad}.} \bibinfo{year}{2020}\natexlab{}.
\newblock \showarticletitle{Opportunities and challenges in explainable artificial intelligence (xai): A survey}.
\newblock \bibinfo{journal}{\emph{arXiv preprint arXiv:2006.11371}} (\bibinfo{year}{2020}).
\newblock


\bibitem[De~Winter and Dodou(2012)]%
        {de2012factor}
\bibfield{author}{\bibinfo{person}{Joost~CF De~Winter} {and} \bibinfo{person}{Dimitra Dodou}.} \bibinfo{year}{2012}\natexlab{}.
\newblock \showarticletitle{Factor recovery by principal axis factoring and maximum likelihood factor analysis as a function of factor pattern and sample size}.
\newblock \bibinfo{journal}{\emph{Journal of applied statistics}} \bibinfo{volume}{39}, \bibinfo{number}{4} (\bibinfo{year}{2012}), \bibinfo{pages}{695--710}.
\newblock


\bibitem[Difallah et~al\mbox{.}(2018)]%
        {difallah2018demographics}
\bibfield{author}{\bibinfo{person}{Djellel Difallah}, \bibinfo{person}{Elena Filatova}, {and} \bibinfo{person}{Panos Ipeirotis}.} \bibinfo{year}{2018}\natexlab{}.
\newblock \showarticletitle{Demographics and dynamics of mechanical turk workers}. In \bibinfo{booktitle}{\emph{Proceedings of the eleventh ACM international conference on web search and data mining}}. \bibinfo{pages}{135--143}.
\newblock


\bibitem[Doshi-Velez and Kim(2017)]%
        {doshi2017towards}
\bibfield{author}{\bibinfo{person}{Finale Doshi-Velez} {and} \bibinfo{person}{Been Kim}.} \bibinfo{year}{2017}\natexlab{}.
\newblock \showarticletitle{Towards a rigorous science of interpretable machine learning}.
\newblock \bibinfo{journal}{\emph{arXiv preprint arXiv:1702.08608}} (\bibinfo{year}{2017}).
\newblock


\bibitem[Doshi-Velez and Kim(2018)]%
        {doshi2018considerations}
\bibfield{author}{\bibinfo{person}{Finale Doshi-Velez} {and} \bibinfo{person}{Been Kim}.} \bibinfo{year}{2018}\natexlab{}.
\newblock \showarticletitle{Considerations for evaluation and generalization in interpretable machine learning}.
\newblock In \bibinfo{booktitle}{\emph{Explainable and interpretable models in computer vision and machine learning}}. \bibinfo{publisher}{Springer}, \bibinfo{pages}{3--17}.
\newblock


\bibitem[Furr(2011)]%
        {furr2011scale}
\bibfield{author}{\bibinfo{person}{Mike Furr}.} \bibinfo{year}{2011}\natexlab{}.
\newblock \bibinfo{booktitle}{\emph{Scale construction and psychometrics for social and personality psychology}}.
\newblock \bibinfo{publisher}{SAGE publications ltd}.
\newblock


\bibitem[Furr(2021)]%
        {furr2021psychometrics}
\bibfield{author}{\bibinfo{person}{R~Michael Furr}.} \bibinfo{year}{2021}\natexlab{}.
\newblock \bibinfo{booktitle}{\emph{Psychometrics: an introduction}}.
\newblock \bibinfo{publisher}{SAGE publications}.
\newblock


\bibitem[Gilpin et~al\mbox{.}(2018)]%
        {gilpin2018explaining}
\bibfield{author}{\bibinfo{person}{Leilani~H Gilpin}, \bibinfo{person}{David Bau}, \bibinfo{person}{Ben~Z Yuan}, \bibinfo{person}{Ayesha Bajwa}, \bibinfo{person}{Michael Specter}, {and} \bibinfo{person}{Lalana Kagal}.} \bibinfo{year}{2018}\natexlab{}.
\newblock \showarticletitle{Explaining explanations: An overview of interpretability of machine learning}. In \bibinfo{booktitle}{\emph{2018 IEEE 5th International Conference on data science and advanced analytics (DSAA)}}. IEEE, \bibinfo{pages}{80--89}.
\newblock


\bibitem[Goldberg and Kilkowski(1985)]%
        {goldberg1985prediction}
\bibfield{author}{\bibinfo{person}{Lewis~R Goldberg} {and} \bibinfo{person}{James~M Kilkowski}.} \bibinfo{year}{1985}\natexlab{}.
\newblock \showarticletitle{The prediction of semantic consistency in self-descriptions: Characteristics of persons and of terms that affect the consistency of responses to synonym and antonym pairs.}
\newblock \bibinfo{journal}{\emph{Journal of personality and social psychology}} \bibinfo{volume}{48}, \bibinfo{number}{1} (\bibinfo{year}{1985}), \bibinfo{pages}{82}.
\newblock


\bibitem[Greisdorf(2003)]%
        {greisdorf2003relevance}
\bibfield{author}{\bibinfo{person}{Howard Greisdorf}.} \bibinfo{year}{2003}\natexlab{}.
\newblock \showarticletitle{Relevance thresholds: a multi-stage predictive model of how users evaluate information}.
\newblock \bibinfo{journal}{\emph{Information Processing \& Management}} \bibinfo{volume}{39}, \bibinfo{number}{3} (\bibinfo{year}{2003}), \bibinfo{pages}{403--423}.
\newblock


\bibitem[Hara et~al\mbox{.}(2018)]%
        {hara2018data}
\bibfield{author}{\bibinfo{person}{Kotaro Hara}, \bibinfo{person}{Abigail Adams}, \bibinfo{person}{Kristy Milland}, \bibinfo{person}{Saiph Savage}, \bibinfo{person}{Chris Callison-Burch}, {and} \bibinfo{person}{Jeffrey~P Bigham}.} \bibinfo{year}{2018}\natexlab{}.
\newblock \showarticletitle{A data-driven analysis of workers' earnings on Amazon Mechanical Turk}. In \bibinfo{booktitle}{\emph{Proceedings of the 2018 CHI conference on human factors in computing systems}}. \bibinfo{pages}{1--14}.
\newblock


\bibitem[Hart(2006)]%
        {hart2006nasa}
\bibfield{author}{\bibinfo{person}{Sandra~G Hart}.} \bibinfo{year}{2006}\natexlab{}.
\newblock \showarticletitle{NASA-task load index (NASA-TLX); 20 years later}. In \bibinfo{booktitle}{\emph{Proceedings of the human factors and ergonomics society annual meeting}}, Vol.~\bibinfo{volume}{50}. Sage publications Sage CA: Los Angeles, CA, \bibinfo{pages}{904--908}.
\newblock


\bibitem[Herman(2017)]%
        {herman2017promise}
\bibfield{author}{\bibinfo{person}{Bernease Herman}.} \bibinfo{year}{2017}\natexlab{}.
\newblock \showarticletitle{The promise and peril of human evaluation for model interpretability}.
\newblock \bibinfo{journal}{\emph{arXiv preprint arXiv:1711.07414}} (\bibinfo{year}{2017}).
\newblock


\bibitem[Hooper et~al\mbox{.}(2008)]%
        {hooper2008structural}
\bibfield{author}{\bibinfo{person}{Daire Hooper}, \bibinfo{person}{Joseph Coughlan}, {and} \bibinfo{person}{Michael~R Mullen}.} \bibinfo{year}{2008}\natexlab{}.
\newblock \showarticletitle{Structural equation modelling: Guidelines for determining model fit}.
\newblock \bibinfo{journal}{\emph{Electronic journal of business research methods}} \bibinfo{volume}{6}, \bibinfo{number}{1} (\bibinfo{year}{2008}), \bibinfo{pages}{pp53--60}.
\newblock


\bibitem[Horn(1965)]%
        {horn1965rationale}
\bibfield{author}{\bibinfo{person}{John~L Horn}.} \bibinfo{year}{1965}\natexlab{}.
\newblock \showarticletitle{A rationale and test for the number of factors in factor analysis}.
\newblock \bibinfo{journal}{\emph{Psychometrika}} \bibinfo{volume}{30}, \bibinfo{number}{2} (\bibinfo{year}{1965}), \bibinfo{pages}{179--185}.
\newblock


\bibitem[Hu and Bentler(1999)]%
        {hu1999cutoff}
\bibfield{author}{\bibinfo{person}{Li-tze Hu} {and} \bibinfo{person}{Peter~M Bentler}.} \bibinfo{year}{1999}\natexlab{}.
\newblock \showarticletitle{Cutoff criteria for fit indexes in covariance structure analysis: Conventional criteria versus new alternatives}.
\newblock \bibinfo{journal}{\emph{Structural equation modeling: a multidisciplinary journal}} \bibinfo{volume}{6}, \bibinfo{number}{1} (\bibinfo{year}{1999}), \bibinfo{pages}{1--55}.
\newblock


\bibitem[Jacovi and Goldberg(2020)]%
        {jacovi2020towards}
\bibfield{author}{\bibinfo{person}{Alon Jacovi} {and} \bibinfo{person}{Yoav Goldberg}.} \bibinfo{year}{2020}\natexlab{}.
\newblock \showarticletitle{Towards faithfully interpretable NLP systems: How should we define and evaluate faithfulness?}
\newblock \bibinfo{journal}{\emph{arXiv preprint arXiv:2004.03685}} (\bibinfo{year}{2020}).
\newblock


\bibitem[Johnson(2005)]%
        {johnson2005ascertaining}
\bibfield{author}{\bibinfo{person}{John~A Johnson}.} \bibinfo{year}{2005}\natexlab{}.
\newblock \showarticletitle{Ascertaining the validity of individual protocols from web-based personality inventories}.
\newblock \bibinfo{journal}{\emph{Journal of research in personality}} \bibinfo{volume}{39}, \bibinfo{number}{1} (\bibinfo{year}{2005}), \bibinfo{pages}{103--129}.
\newblock


\bibitem[Kaiser(1960)]%
        {kaiser1960application}
\bibfield{author}{\bibinfo{person}{Henry~F Kaiser}.} \bibinfo{year}{1960}\natexlab{}.
\newblock \showarticletitle{The application of electronic computers to factor analysis}.
\newblock \bibinfo{journal}{\emph{Educational and psychological measurement}} \bibinfo{volume}{20}, \bibinfo{number}{1} (\bibinfo{year}{1960}), \bibinfo{pages}{141--151}.
\newblock


\bibitem[Kim et~al\mbox{.}(2016)]%
        {kim2016examples}
\bibfield{author}{\bibinfo{person}{Been Kim}, \bibinfo{person}{Rajiv Khanna}, {and} \bibinfo{person}{Oluwasanmi~O Koyejo}.} \bibinfo{year}{2016}\natexlab{}.
\newblock \showarticletitle{Examples are not enough, learn to criticize! criticism for interpretability}.
\newblock \bibinfo{journal}{\emph{Advances in neural information processing systems}}  \bibinfo{volume}{29} (\bibinfo{year}{2016}).
\newblock


\bibitem[Kim et~al\mbox{.}(2022)]%
        {kim2022understanding}
\bibfield{author}{\bibinfo{person}{Hyeji Kim}, \bibinfo{person}{Inchan Jung}, {and} \bibinfo{person}{Youn-kyung Lim}.} \bibinfo{year}{2022}\natexlab{}.
\newblock \showarticletitle{Understanding the Negative Aspects of User Experience in Human-likeness of Voice-based Conversational Agents}. In \bibinfo{booktitle}{\emph{Designing Interactive Systems Conference}}. \bibinfo{pages}{1418--1427}.
\newblock


\bibitem[Kline(2005)]%
        {kline2005psychological}
\bibfield{author}{\bibinfo{person}{Theresa Kline}.} \bibinfo{year}{2005}\natexlab{}.
\newblock \bibinfo{booktitle}{\emph{Psychological testing: A practical approach to design and evaluation}}.
\newblock \bibinfo{publisher}{Sage}.
\newblock


\bibitem[Lage et~al\mbox{.}(2019)]%
        {lage2019human}
\bibfield{author}{\bibinfo{person}{Isaac Lage}, \bibinfo{person}{Emily Chen}, \bibinfo{person}{Jeffrey He}, \bibinfo{person}{Menaka Narayanan}, \bibinfo{person}{Been Kim}, \bibinfo{person}{Samuel~J Gershman}, {and} \bibinfo{person}{Finale Doshi-Velez}.} \bibinfo{year}{2019}\natexlab{}.
\newblock \showarticletitle{Human evaluation of models built for interpretability}. In \bibinfo{booktitle}{\emph{Proceedings of the AAAI Conference on Human Computation and Crowdsourcing}}, Vol.~\bibinfo{volume}{7}. \bibinfo{pages}{59--67}.
\newblock


\bibitem[Lipton(2018)]%
        {lipton2018mythos}
\bibfield{author}{\bibinfo{person}{Zachary~C Lipton}.} \bibinfo{year}{2018}\natexlab{}.
\newblock \showarticletitle{The Mythos of Model Interpretability: In machine learning, the concept of interpretability is both important and slippery.}
\newblock \bibinfo{journal}{\emph{Queue}} \bibinfo{volume}{16}, \bibinfo{number}{3} (\bibinfo{year}{2018}), \bibinfo{pages}{31--57}.
\newblock


\bibitem[Lundberg and Lee(2017)]%
        {lundberg2017unified}
\bibfield{author}{\bibinfo{person}{Scott~M Lundberg} {and} \bibinfo{person}{Su-In Lee}.} \bibinfo{year}{2017}\natexlab{}.
\newblock \showarticletitle{A unified approach to interpreting model predictions}.
\newblock \bibinfo{journal}{\emph{Advances in neural information processing systems}}  \bibinfo{volume}{30} (\bibinfo{year}{2017}).
\newblock


\bibitem[MacCallum et~al\mbox{.}(1999)]%
        {maccallum1999sample}
\bibfield{author}{\bibinfo{person}{Robert~C MacCallum}, \bibinfo{person}{Keith~F Widaman}, \bibinfo{person}{Shaobo Zhang}, {and} \bibinfo{person}{Sehee Hong}.} \bibinfo{year}{1999}\natexlab{}.
\newblock \showarticletitle{Sample size in factor analysis.}
\newblock \bibinfo{journal}{\emph{Psychological methods}} \bibinfo{volume}{4}, \bibinfo{number}{1} (\bibinfo{year}{1999}), \bibinfo{pages}{84}.
\newblock


\bibitem[Meade and Craig(2012)]%
        {meade2012identifying}
\bibfield{author}{\bibinfo{person}{Adam~W Meade} {and} \bibinfo{person}{S~Bartholomew Craig}.} \bibinfo{year}{2012}\natexlab{}.
\newblock \showarticletitle{Identifying careless responses in survey data.}
\newblock \bibinfo{journal}{\emph{Psychological methods}} \bibinfo{volume}{17}, \bibinfo{number}{3} (\bibinfo{year}{2012}), \bibinfo{pages}{437}.
\newblock


\bibitem[Nauta et~al\mbox{.}(2022)]%
        {nauta2022anecdotal}
\bibfield{author}{\bibinfo{person}{Meike Nauta}, \bibinfo{person}{Jan Trienes}, \bibinfo{person}{Shreyasi Pathak}, \bibinfo{person}{Elisa Nguyen}, \bibinfo{person}{Michelle Peters}, \bibinfo{person}{Yasmin Schmitt}, \bibinfo{person}{J{\"o}rg Schl{\"o}tterer}, \bibinfo{person}{Maurice van Keulen}, {and} \bibinfo{person}{Christin Seifert}.} \bibinfo{year}{2022}\natexlab{}.
\newblock \showarticletitle{From anecdotal evidence to quantitative evaluation methods: A systematic review on evaluating explainable ai}.
\newblock \bibinfo{journal}{\emph{arXiv preprint arXiv:2201.08164}} (\bibinfo{year}{2022}).
\newblock


\bibitem[Nguyen and Mart{\'\i}nez(2020)]%
        {nguyen2020quantitative}
\bibfield{author}{\bibinfo{person}{An-phi Nguyen} {and} \bibinfo{person}{Mar{\'\i}a~Rodr{\'\i}guez Mart{\'\i}nez}.} \bibinfo{year}{2020}\natexlab{}.
\newblock \showarticletitle{On quantitative aspects of model interpretability}.
\newblock \bibinfo{journal}{\emph{arXiv preprint arXiv:2007.07584}} (\bibinfo{year}{2020}).
\newblock


\bibitem[Noman et~al\mbox{.}(2019)]%
        {noman2019techies}
\bibfield{author}{\bibinfo{person}{Abu Saleh~Md Noman}, \bibinfo{person}{Sanchari Das}, {and} \bibinfo{person}{Sameer Patil}.} \bibinfo{year}{2019}\natexlab{}.
\newblock \showarticletitle{Techies against Facebook: understanding negative sentiment toward Facebook via user generated content}. In \bibinfo{booktitle}{\emph{Proceedings of the 2019 CHI Conference on Human Factors in Computing Systems}}. \bibinfo{pages}{1--15}.
\newblock


\bibitem[Olmsted-Hawala et~al\mbox{.}(2010)]%
        {olmsted2010think}
\bibfield{author}{\bibinfo{person}{Erica~L Olmsted-Hawala}, \bibinfo{person}{Elizabeth~D Murphy}, \bibinfo{person}{Sam Hawala}, {and} \bibinfo{person}{Kathleen~T Ashenfelter}.} \bibinfo{year}{2010}\natexlab{}.
\newblock \showarticletitle{Think-aloud protocols: a comparison of three think-aloud protocols for use in testing data-dissemination web sites for usability}. In \bibinfo{booktitle}{\emph{Proceedings of the SIGCHI conference on human factors in computing systems}}. \bibinfo{pages}{2381--2390}.
\newblock


\bibitem[Peer et~al\mbox{.}(2014)]%
        {peer2014reputation}
\bibfield{author}{\bibinfo{person}{Eyal Peer}, \bibinfo{person}{Joachim Vosgerau}, {and} \bibinfo{person}{Alessandro Acquisti}.} \bibinfo{year}{2014}\natexlab{}.
\newblock \showarticletitle{Reputation as a sufficient condition for data quality on Amazon Mechanical Turk}.
\newblock \bibinfo{journal}{\emph{Behavior research methods}} \bibinfo{volume}{46}, \bibinfo{number}{4} (\bibinfo{year}{2014}), \bibinfo{pages}{1023--1031}.
\newblock


\bibitem[Polley et~al\mbox{.}(2021)]%
        {polley2021towards}
\bibfield{author}{\bibinfo{person}{Sayantan Polley}, \bibinfo{person}{Rashmi~Raju Koparde}, \bibinfo{person}{Akshaya~Bindu Gowri}, \bibinfo{person}{Maneendra Perera}, {and} \bibinfo{person}{Andreas Nuernberger}.} \bibinfo{year}{2021}\natexlab{}.
\newblock \showarticletitle{Towards trustworthiness in the context of explainable search}. In \bibinfo{booktitle}{\emph{Proceedings of the 44th International ACM SIGIR Conference on Research and Development in Information Retrieval}}. \bibinfo{pages}{2580--2584}.
\newblock


\bibitem[Qu et~al\mbox{.}(2020)]%
        {qu2020towards}
\bibfield{author}{\bibinfo{person}{Jiaming Qu}, \bibinfo{person}{Jaime Arguello}, {and} \bibinfo{person}{Yue Wang}.} \bibinfo{year}{2020}\natexlab{}.
\newblock \showarticletitle{Towards explainable retrieval models for precision medicine literature search}. In \bibinfo{booktitle}{\emph{Proceedings of the 43rd International ACM SIGIR Conference on Research and Development in Information Retrieval}}. \bibinfo{pages}{1593--1596}.
\newblock


\bibitem[Qvarfordt et~al\mbox{.}(2013)]%
        {qvarfordt2013looking}
\bibfield{author}{\bibinfo{person}{Pernilla Qvarfordt}, \bibinfo{person}{Gene Golovchinsky}, \bibinfo{person}{Tony Dunnigan}, {and} \bibinfo{person}{Elena Agapie}.} \bibinfo{year}{2013}\natexlab{}.
\newblock \showarticletitle{Looking ahead: Query preview in exploratory search}. In \bibinfo{booktitle}{\emph{Proceedings of the 36th international ACM SIGIR conference on Research and development in information retrieval}}. \bibinfo{pages}{243--252}.
\newblock


\bibitem[Ramos and Eickhoff(2020)]%
        {ramos2020search}
\bibfield{author}{\bibinfo{person}{Jerome Ramos} {and} \bibinfo{person}{Carsten Eickhoff}.} \bibinfo{year}{2020}\natexlab{}.
\newblock \showarticletitle{Search result explanations improve efficiency and trust}. In \bibinfo{booktitle}{\emph{Proceedings of the 43rd International ACM SIGIR Conference on Research and Development in Information Retrieval}}. \bibinfo{pages}{1597--1600}.
\newblock


\bibitem[Ribeiro et~al\mbox{.}(2016)]%
        {ribeiro2016should}
\bibfield{author}{\bibinfo{person}{Marco~Tulio Ribeiro}, \bibinfo{person}{Sameer Singh}, {and} \bibinfo{person}{Carlos Guestrin}.} \bibinfo{year}{2016}\natexlab{}.
\newblock \showarticletitle{" Why should i trust you?" Explaining the predictions of any classifier}. In \bibinfo{booktitle}{\emph{Proceedings of the 22nd ACM SIGKDD international conference on knowledge discovery and data mining}}. \bibinfo{pages}{1135--1144}.
\newblock


\bibitem[Schnabel et~al\mbox{.}(2020)]%
        {schnabel2020impact}
\bibfield{author}{\bibinfo{person}{Tobias Schnabel}, \bibinfo{person}{Saleema Amershi}, \bibinfo{person}{Paul~N Bennett}, \bibinfo{person}{Peter Bailey}, {and} \bibinfo{person}{Thorsten Joachims}.} \bibinfo{year}{2020}\natexlab{}.
\newblock \showarticletitle{The Impact of More Transparent Interfaces on Behavior in Personalized Recommendation}. In \bibinfo{booktitle}{\emph{Proceedings of the 43rd International ACM SIGIR Conference on Research and Development in Information Retrieval}}. \bibinfo{pages}{991--1000}.
\newblock


\bibitem[Singh and Anand(2019)]%
        {singh2019exs}
\bibfield{author}{\bibinfo{person}{Jaspreet Singh} {and} \bibinfo{person}{Avishek Anand}.} \bibinfo{year}{2019}\natexlab{}.
\newblock \showarticletitle{Exs: Explainable search using local model agnostic interpretability}. In \bibinfo{booktitle}{\emph{Proceedings of the Twelfth ACM International Conference on Web Search and Data Mining}}. \bibinfo{pages}{770--773}.
\newblock


\bibitem[Slack et~al\mbox{.}(2019)]%
        {slack2019assessing}
\bibfield{author}{\bibinfo{person}{Dylan Slack}, \bibinfo{person}{Sorelle~A Friedler}, \bibinfo{person}{Carlos Scheidegger}, {and} \bibinfo{person}{Chitradeep~Dutta Roy}.} \bibinfo{year}{2019}\natexlab{}.
\newblock \showarticletitle{Assessing the local interpretability of machine learning models}.
\newblock \bibinfo{journal}{\emph{arXiv preprint arXiv:1902.03501}} (\bibinfo{year}{2019}).
\newblock


\bibitem[Steiger(2007)]%
        {steiger2007understanding}
\bibfield{author}{\bibinfo{person}{James~H Steiger}.} \bibinfo{year}{2007}\natexlab{}.
\newblock \showarticletitle{Understanding the limitations of global fit assessment in structural equation modeling}.
\newblock \bibinfo{journal}{\emph{Personality and Individual differences}} \bibinfo{volume}{42}, \bibinfo{number}{5} (\bibinfo{year}{2007}), \bibinfo{pages}{893--898}.
\newblock


\bibitem[Tabachnick et~al\mbox{.}(2007)]%
        {tabachnick2007using}
\bibfield{author}{\bibinfo{person}{Barbara~G Tabachnick}, \bibinfo{person}{Linda~S Fidell}, {and} \bibinfo{person}{Jodie~B Ullman}.} \bibinfo{year}{2007}\natexlab{}.
\newblock \bibinfo{booktitle}{\emph{Using multivariate statistics}}. Vol.~\bibinfo{volume}{5}.
\newblock \bibinfo{publisher}{pearson Boston, MA}.
\newblock


\bibitem[Thurstone(1947)]%
        {thurstone1947multiple}
\bibfield{author}{\bibinfo{person}{Louis~Leon Thurstone}.} \bibinfo{year}{1947}\natexlab{}.
\newblock \showarticletitle{Multiple-factor analysis; a development and expansion of The Vectors of Mind.}
\newblock  (\bibinfo{year}{1947}).
\newblock


\bibitem[Ullman and Bentler(2012)]%
        {ullman2012structural}
\bibfield{author}{\bibinfo{person}{Jodie~B Ullman} {and} \bibinfo{person}{Peter~M Bentler}.} \bibinfo{year}{2012}\natexlab{}.
\newblock \showarticletitle{Structural equation modeling}.
\newblock \bibinfo{journal}{\emph{Handbook of Psychology, Second Edition}}  \bibinfo{volume}{2} (\bibinfo{year}{2012}).
\newblock


\bibitem[Voorhees(2005)]%
        {voorhees_overview_2005}
\bibfield{author}{\bibinfo{person}{Ellen Voorhees}.} \bibinfo{year}{2005}\natexlab{}.
\newblock \bibinfo{title}{Overview of the {TREC} 2004 {Robust} {Retrieval} {Track}}.
\newblock
\newblock
\urldef\tempurl%
\url{https://doi.org/10.6028/NIST.SP.500-261}
\showDOI{\tempurl}


\bibitem[Wheaton et~al\mbox{.}(1977)]%
        {wheaton1977assessing}
\bibfield{author}{\bibinfo{person}{Blair Wheaton}, \bibinfo{person}{Bengt Muthen}, \bibinfo{person}{Duane~F Alwin}, {and} \bibinfo{person}{Gene~F Summers}.} \bibinfo{year}{1977}\natexlab{}.
\newblock \showarticletitle{Assessing reliability and stability in panel models}.
\newblock \bibinfo{journal}{\emph{Sociological methodology}}  \bibinfo{volume}{8} (\bibinfo{year}{1977}), \bibinfo{pages}{84--136}.
\newblock


\bibitem[Williams et~al\mbox{.}(2010)]%
        {williams2010exploratory}
\bibfield{author}{\bibinfo{person}{Brett Williams}, \bibinfo{person}{Andrys Onsman}, {and} \bibinfo{person}{Ted Brown}.} \bibinfo{year}{2010}\natexlab{}.
\newblock \showarticletitle{Exploratory factor analysis: A five-step guide for novices}.
\newblock \bibinfo{journal}{\emph{Australasian journal of paramedicine}} \bibinfo{volume}{8}, \bibinfo{number}{3} (\bibinfo{year}{2010}).
\newblock


\bibitem[Worthington and Whittaker(2006)]%
        {worthington2006scale}
\bibfield{author}{\bibinfo{person}{Roger~L Worthington} {and} \bibinfo{person}{Tiffany~A Whittaker}.} \bibinfo{year}{2006}\natexlab{}.
\newblock \showarticletitle{Scale development research: A content analysis and recommendations for best practices}.
\newblock \bibinfo{journal}{\emph{The counseling psychologist}} \bibinfo{volume}{34}, \bibinfo{number}{6} (\bibinfo{year}{2006}), \bibinfo{pages}{806--838}.
\newblock


\bibitem[Yu et~al\mbox{.}(2022)]%
        {yu2022towards}
\bibfield{author}{\bibinfo{person}{Puxuan Yu}, \bibinfo{person}{Razieh Rahimi}, {and} \bibinfo{person}{James Allan}.} \bibinfo{year}{2022}\natexlab{}.
\newblock \showarticletitle{Towards Explainable Search Results: A Listwise Explanation Generator}. In \bibinfo{booktitle}{\emph{Proceedings of the 45th International ACM SIGIR Conference on Research and Development in Information Retrieval}}. \bibinfo{pages}{669--680}.
\newblock


\bibitem[Zhang et~al\mbox{.}(2014)]%
        {zhang2014multidimensional}
\bibfield{author}{\bibinfo{person}{Yinglong Zhang}, \bibinfo{person}{Jin Zhang}, \bibinfo{person}{Matthew Lease}, {and} \bibinfo{person}{Jacek Gwizdka}.} \bibinfo{year}{2014}\natexlab{}.
\newblock \showarticletitle{Multidimensional relevance modeling via psychometrics and crowdsourcing}. In \bibinfo{booktitle}{\emph{Proceedings of the 37th international ACM SIGIR conference on Research \& development in information retrieval}}. \bibinfo{pages}{435--444}.
\newblock


\end{thebibliography}

\end{document}